\documentclass{article}

\usepackage{arxiv}

\usepackage[utf8]{inputenc} 
\usepackage[T1]{fontenc}    
\usepackage{hyperref}       
\usepackage{url}            
\usepackage{booktabs}       
\usepackage{amsfonts}       
\usepackage{nicefrac}       
\usepackage{microtype}      
\usepackage{lipsum}		
\usepackage{graphicx}
\usepackage{natbib}
\usepackage{doi}
\usepackage{amsmath}
\usepackage{amssymb}

\title{Evaluation of Transverse Incoherent Tune Shifts in the Electron-Ion Collider}

\author{ \href{https://orcid.org/0000-0002-5560-6419}{\includegraphics[scale=0.06]{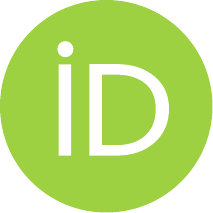}\hspace{1mm}Xiaofeng Gu}\thanks{Use footnote for providing further
		information about author (webpage, alternative
		address)---\emph{not} for acknowledging funding agencies.} \\
	Accelerator Divsion\\
	Brookhaven National Laboratory\\
	Uption, NY 11973, USA \\
	\texttt{xgu@bnl.gov} \\
	\And
	\href{https://orcid.org/0000-0001-9391-1820}{\includegraphics[scale=0.06]{orcid.pdf}\hspace{1mm}Alexei Blednykh} \\
	Accelerator Divsion\\
	Brookhaven National Laboratory\\
	Uption, NY 11973, USA \\
}

\renewcommand{\undertitle}{Technical Report}

\hypersetup{
}

\setcitestyle{numbers}

\begin{document}
\maketitle





\begin{abstract}

To achieve the peak luminosity goals ($10^{34}\,\mathrm{cm}^{-2}\mathrm{s}^{-1}$) of the Electron-Ion Collider (EIC), the Hadron Storage Ring (HSR) and Electron Storage Ring (ESR) must operate with high average currents, flat beam geometries, non-circular vacuum chambers, and large transverse orbit offsets. These conditions can produce strong collective effects and incoherent tune shifts whose associated tune spreads can push individual particle tunes across low-order betatron resonance lines~\cite{Peggs2026_EICPDR}, leading to slow emittance growth, halo formation, and reduced beam lifetime.

In this work, we present a comprehensive quantitative evaluation of incoherent transverse betatron tune shifts for the EIC. The analysis includes three contributions: beam–beam interactions, direct space-charge effects, and beam–environment interactions, the latter representing indirect space-charge effects arising from image forces and vacuum chamber geometry. Numerical simulations performed using Xsuite~\cite{Iadarola2024Xsuite} and PyHEADTAIL~\cite{pyheadtail} validate the analytical evaluations for beam–beam and direct space-charge effects, respectively. For the incoherent tune shift induced by indirect space-charge, several formalisms are reviewed and compared. Specifically, we employ the Laslett image-charge formulation for rectangular vacuum chamber geometries, which is applicable to off-axis beam trajectories. This off-axis approach accounts for transverse beam offsets of up to 20 mm expected in the HSR and 10 mm in the ESR.

The results provide fundamental beam parameters and quantitative guidance for the design, optimization, and operational considerations of the EIC. In addition, the overview and comparison of different formalisms for indirect space-charge offer a clear, accessible guide with established baseline parameters and methodology for researchers and newcomers entering high-intensity beam dynamics.

\end{abstract}


\section{Introduction}

\begin{figure}[!ht]
\centering
\includegraphics[width=\columnwidth]{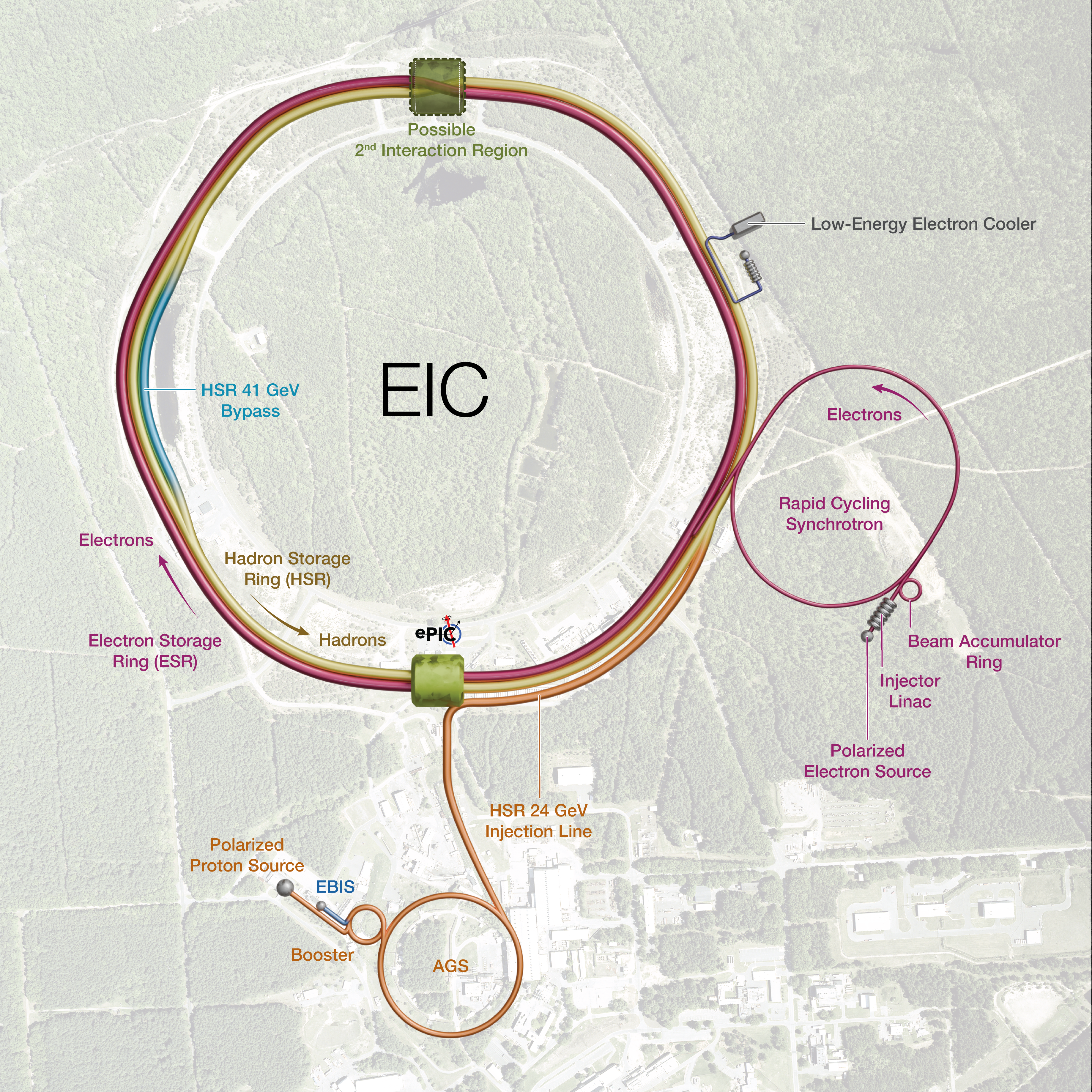}
\caption{Schematic layout of the Electron-Ion Collider facility, illustrating the Hadron Storage Ring (HSR), Electron Storage Ring (ESR), Rapid Cycling Synchrotron (RCS), hadron pre-accelerator complex (EBIS, Booster, AGS), the ePIC interaction region, and a potential second interaction region. This schematic view of the Electron-Ion Collider (EIC) is adapted from the EIC Preliminary Design Report\cite{Peggs2026_EICPDR}.}
\label{figs:EIC_Layout}
\centering
\end{figure}

The Electron-Ion Collider (EIC) is planned for construction at Brookhaven National Laboratory and consists of two main storage rings: the Electron Storage Ring (ESR) and the Hadron Storage Ring (HSR)~\cite{Peggs2026_EICPDR}, as shown in Fig.~\ref{figs:EIC_Layout}. The ESR circulates an intense beam of electron bunches with energies ranging from 5 to 18 GeV, while the HSR stores a high-energy beam of protons or heavier atomic nuclei with collision energies ranging from 41 GeV to 275 GeV.

To achieve high luminosity, several techniques are implemented during the design phase, including the use of flat beams with a horizontal-to-vertical emittance ratio of 10~\cite{Luo2026_EIC_LargeEmittanceRatio}. To match the ESR revolution frequency, a maximum beam radial shift of up to 20 mm is required in the HSR lattice arcs for 100 GeV ion beams~\cite{ROBERTDEMOLAIZE2025170043}. Meanwhile, to achieve the design luminosity, maximum beam currents are set to 2.5 A for the ESR and 1.0 A for the HSR.

\begin{figure*}[!ht]
\centering
\includegraphics[width=\textwidth]{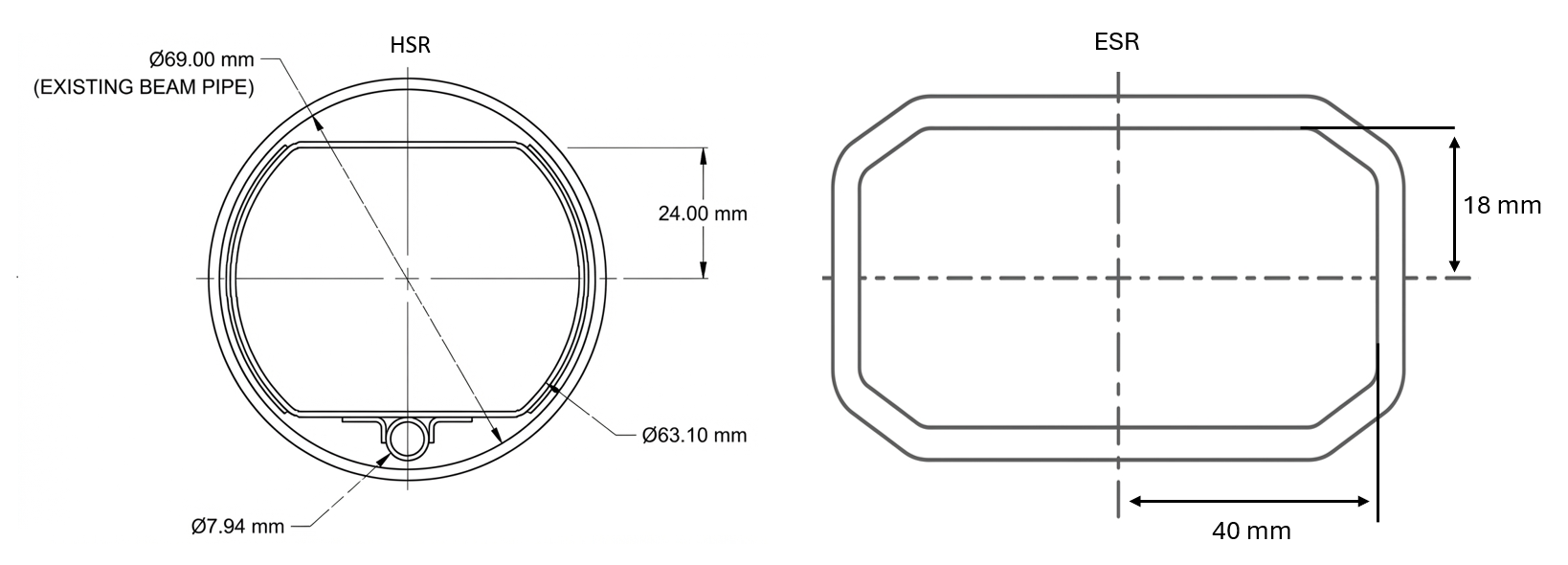}
\caption{Cross-sectional schematic of the Hadron Storage Ring (HSR) and ESR beam screen.}
\label{figs:HSR_Screen}
\centering
\end{figure*}

To accommodate the flat beam shape, a rectangular beam screen or vacuum chamber was adopted for the HSR and ESR. Fig.~\ref{figs:HSR_Screen} depicts the cross-sectional profile of the HSR beam screens and ESR vacuum chamber. The proposed HSR vacuum chamber uses the existing RHIC stainless-steel (316LN)~\cite{Anerella2003_RHIC_magnets} round beam pipe as the outer vacuum enclosure, whose magnetic permeability $\mu$ is extremely close to 1.0 (virtually non-magnetic)~\cite{Wilson1991_magnetic_permeability}, ensuring the beam tube does not distort the precision magnetic fields generated by the surrounding superconducting magnets. On the inner side, a rectangular stainless-steel beam screen coated with a copper base layer and an amorphous-carbon (a-C) top layer will be inserted into the existing circular beam pipe. Meanwhile, the ESR vacuum chamber is constructed from copper with an internal NEG coating.

Given these combined constraints—including high beam current, flat beam cross-sections, rectangular vacuum chamber or beam screen, and transverse beam offsets—the incoherent tune shifts (and resulting tune spread) will be driven by three primary mechanisms: beam-beam collisions, direct space-charge fields, and indirect space-charge fields generated by beam interactions with the surrounding vacuum chamber or beam screen. Therefore, an accurate assessment of incoherent tune shifts is essential for HSR and ESR design, such as working-point optimization, beam stability analysis, and the development of compensation strategies. An excessive tune spread can drive particles across betatron resonances~\cite{Peggs2026_EICPDR}, causing resonance diffusion, emittance growth, halo formation, reduced dynamic aperture, and eventual particle loss.

Regarding the third mechanism, while individual aspects of indirect space-charge and impedance-driven tune shifts are well established in the literature, this work fills a key gap by bridging classical static image-charge formulations (L. Laslett~\cite{Laslett1963}) with modern multi-bunch resistive-wall frameworks (e.g., A. Chao~\cite{ChaoHeifetsZotter2002}, A. Blednykh~\cite{Blednykh2016_LowFrequencyQuadrupoleImpedance}, and Y. Shobuda~\cite{ShobudaYokoya2002_ResistiveWall}). Applying these models to the ESR across multiple operational energies ($5$, $10$, and $18~\mathrm{GeV}$) enables direct cross-validation of the analytical calculations. Ultimately, this comprehensive overview and comparison provides a clear, accessible reference with established baseline parameters and methodologies for researchers and newcomers in high-intensity beam dynamics.

The remainder of this paper is organized as follows. Section~\ref{sec: incoherent tune shift} introduces the three principal sources of incoherent tune shifts and presents the beam--beam and direct space-charge evaluations. The indirect space-charge-induced quadrupolar tune shifts are then reviewed and compared using the A.~Chao, A.~Blednykh, Y.~Shobuda, and L.~Laslett formalisms, with guidance on their applicability to EIC chamber geometries and off-axis trajectories. The subsequent section~\ref{sec: ShortWakeField} describes the evaluation of single-bunch short-range wakefield effects from electromagnetic simulations. Finally, the last section~\ref{sec: Discussion} summarizes the implications for EIC working-point control, tune feedback, and future collective-effects studies.

\section{\label{sec: incoherent tune shift} Incoherent Tune shifts}

\subsection{\label{sec: beam-beam} Beam–Beam Incoherent Tune Shift}

During collisions, charged particles in a bunch experience strong electromagnetic fields produced by the opposing beam. The beam–beam parameter ($\xi$) characterizes the overall strength of this interaction. For Gaussian beams in head-on collisions, the beam–beam parameter is expressed as~\cite{Peggs2026_EICPDR, Ohmi2024_SpaceChargeBeamBeam}:

\begin{equation}
\xi_{e,p,z}^{bb}=\frac{ r_{e,p}\beta^{*}_{e,p,z}}{2\pi\gamma_{e,p}}\,\frac{N_{{p,e}}}{\sigma^{*}_{{p,e},z} 
\left(\sigma^{*}_{{{p,e}},x}+\sigma^{*}_{{{p,e}},y}\right)}
\label{eq:beam_beam_parameter}
\end{equation}
where $z \in \{x, y\}$ denotes the horizontal or vertical plane, $r_{e,p}$ is the classical radius of the electron or proton, $\gamma_{e,p}$ is the relativistic Lorentz factor of the corresponding beam, and $\beta^{*}_{e,p,z}$ is the beta function at the interaction point (IP). The parameters $N_{p,e}$ and $\sigma^{*}_{p,e,z}$ represent the bunch intensity and transverse RMS beam size of the opposing beam (proton or electron) at the IP, respectively. Thus, when evaluating the beam-beam parameter of the proton beam, the electron bunch intensity and transverse beam size are used, and vice versa.

\begin{figure}[!ht]
\centering
\includegraphics[width=\columnwidth]{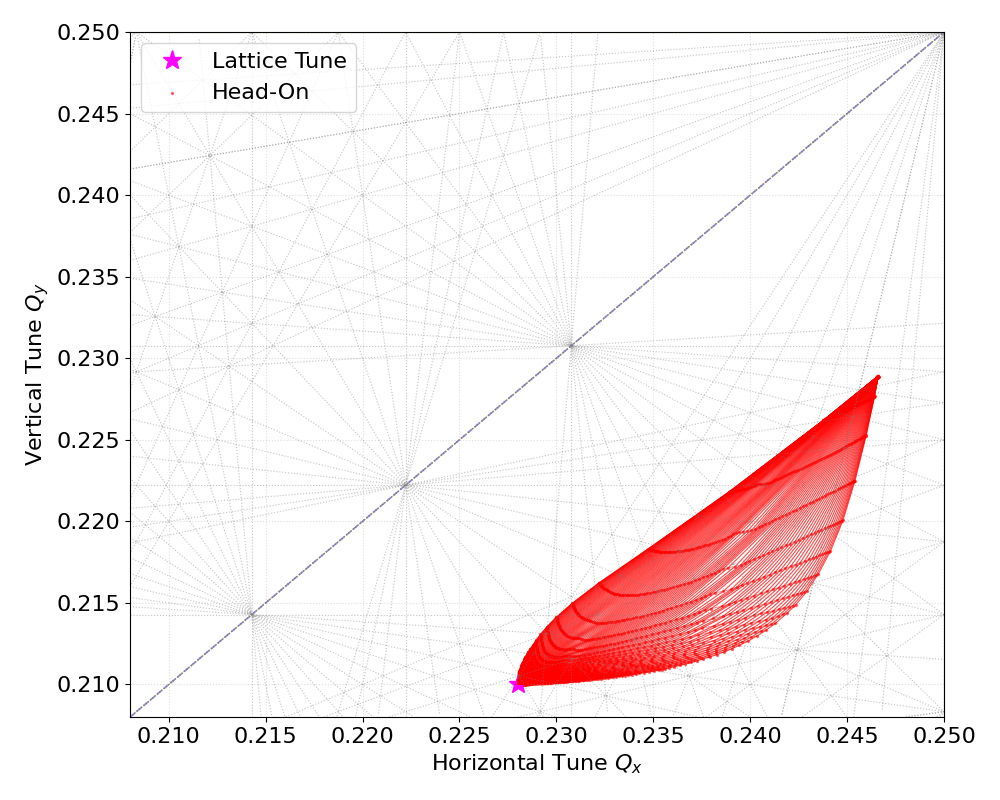}
\caption{The incoherent beam–beam tune shift for the $100\text{ GeV}$ HSR proton beam was evaluated via particle tracking using Xsuite, yielding $\Delta \nu_{x, \text{incoh}}^{\text{bb}} = 0.0186$ and $\Delta \nu_{y, \text{incoh}}^{\text{bb}} = 0.0189$.}
\label{figs:HSR_BeamBeam}
\centering
\end{figure}

For particles executing small-amplitude betatron oscillations near the bunch core, the beam–beam force varies linearly with particle offset. With this small-amplitude assumption, the incoherent beam–beam tune shift ($\Delta \nu_{xy, incoh}^{bb}$) equals the beam–beam parameter:
\begin{equation}
\Delta \nu_{xy, incoh}^{bb}  \approx \xi_{xy}
\label{eq:beam_beam_parameter2}
\end{equation}

However, as a particle's oscillation amplitude grows larger than the RMS beam size ($\sigma$), the strong non-linearity of the beam–beam electromagnetic field becomes dominant. Particles executing large-amplitude orbits experience a progressively weaker integrated kick, causing their incoherent tune shifts to decay toward zero. 

This amplitude dependence produces a broad distribution of individual betatron frequencies across the bunch—forming the characteristic beam–beam tune footprint on the working diagram—where the linear parameter $\xi$ defines the maximum tune shift at zero amplitude. 

As an example, Fig.~\ref{figs:HSR_BeamBeam} illustrates the beam–beam tune footprint for the $100\text{ GeV}$ Hadron Storage Ring (HSR) proton lattice. The horizontal and vertical \(\beta^*\) values for the HSR proton lattice are $0.63 \text{m}$ and $0.057 \text{m}$, while the electron bunch intensity is $17.2\cdot 10^{10}$ and its horizontal and vertical electron beam sizes are $110.45\ \mu\text{m}$ and $9.61\ \mu\text{m}$, respectively.

This yields incoherent tune shifts $\Delta \nu_{x, \text{incoh}}^{\text{bb}} = 0.0186$ and $\Delta \nu_{y, \text{incoh}}^{\text{bb}} = 0.0189$. These numerical results show excellent agreement with the theoretical value of $\Delta \nu_{x,y; \text{incoh}}^{\text{bb}} = 0.019$ calculated from analytical formulas in Eq.~\eqref{eq:beam_beam_parameter}.

For the beam-beam parameter evaluation, the bare-lattice (before beam-beam focusing) $\beta^*$ values are used, and no focused or equilibrium beta functions are used in this comparison. All electron and proton beam sizes are evaluated from their existing designed $\beta^*$ and emittance since they are measurable parameters, not their required beam sizes. 

\subsection{\label{sec: Direct space-charge} Direct space-charge incoherent tune shift}

In addition to the incoherent beam–beam tune shift generated by the opposing bunch, moving charged particles also generate electromagnetic self-fields that perturb other particles within the same bunch. This self-interaction causes a transverse defocusing effect known as the space-charge tune shift~\cite{Schindl2006_SpaceCharge}. 

Space-charge effects are generally classified into two categories: direct and indirect. Direct space charge arises from the Coulomb repulsion and magnetic attraction among particles within the beam itself and is discussed in this section. In contrast, indirect space-charge is discussed in Sec.~\ref{sec:indirect-space-charge}.

\begin{figure}[!ht]
\centering
\includegraphics[width=\columnwidth]{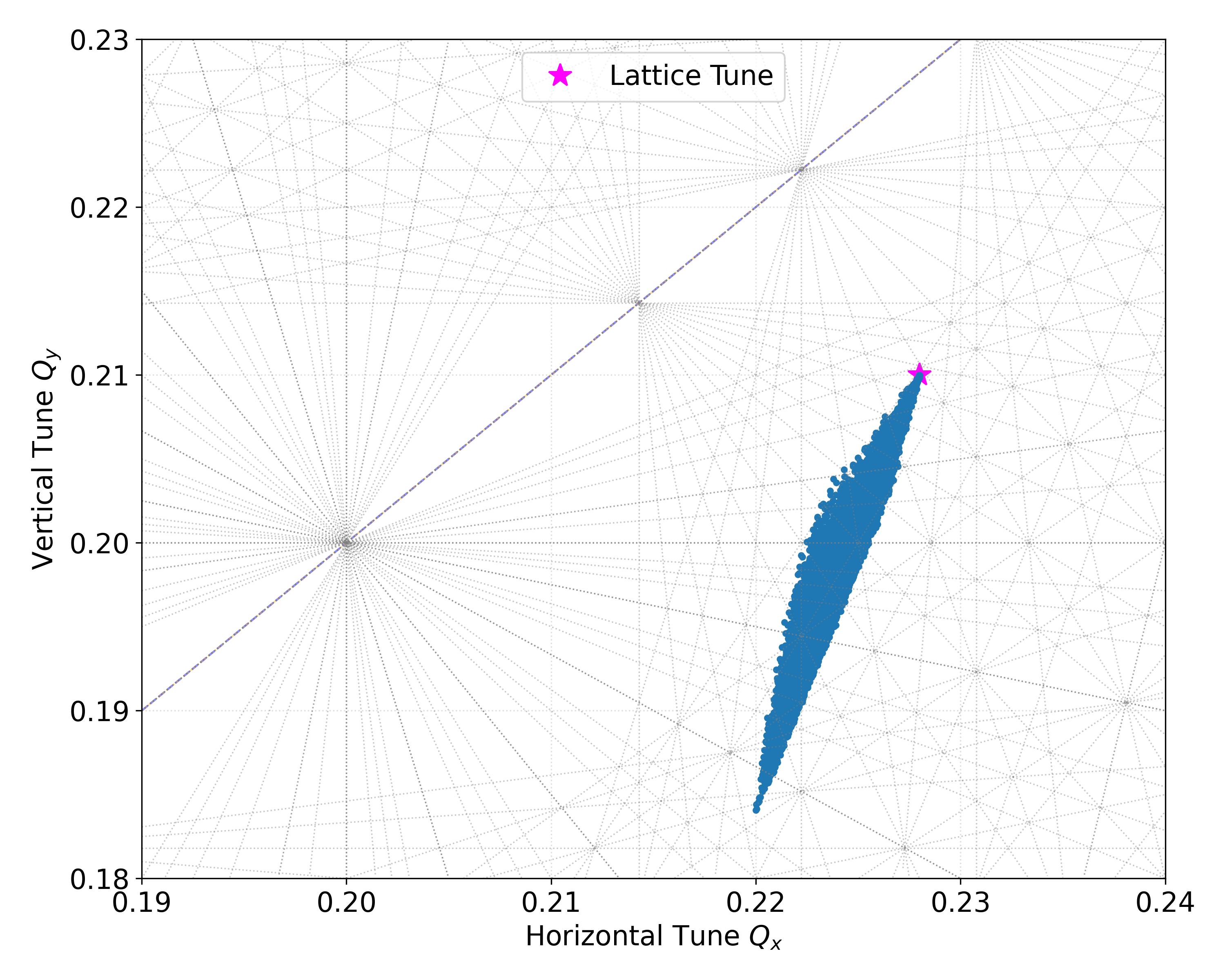}
\caption{The incoherent space-charge tune shift and corresponding tune footprint for the $100\text{ GeV}$ HSR proton beam were evaluated via multi-particle tracking in PyHEADTAIL using a one-turn map based on average beta functions.}
\label{figs:HSR_SpaceCharge}
\centering
\end{figure}

The direct space-charge tune shift for a bunched, bi-Gaussian elliptic beam used in this paper can be expressed as \cite{Oeftiger2022_SpaceChargeLimit, Zampetakis2023_CLIC_DR}:
\begin{equation}
\label{eq:direct-space-charge-gaussian}
\begin{aligned}
\Delta \nu_{xy, incoh}^{sc, direct} &= -\frac{r_0 N_{bunch}}{(2\pi)^{3/2} \sigma_z \beta^2 \gamma^3} \oint ds \frac{\beta_{xy}(s)}{\left[\sigma_x(s) + \sigma_y(s)\right] \sigma_{xy}(s)} 
\end{aligned}
\end{equation}
where $\beta_{x,y}(s)$ and $\sigma_{x,y}(s)$ represent the betatron function and RMS beam size in the horizontal ($x$) or vertical ($y$) plane as functions of the longitudinal position $s$. While the transverse beam sizes are given by:
\begin{equation}
\label{eq:beam_size}
\sigma_{x,y}(s) = \sqrt{\epsilon_{x,y;\text{geom}}\beta_{x,y}(s) + \left[D_{x,y}(s) \cdot \delta_p\right]^2},
\end{equation}
with $\epsilon_{x,y;\text{geom}}$ denoting the geometric emittances and $D_{x,y}(s)$ representing the dispersion functions. For planar lattices without vertical dispersion, $D_y(s) = 0$.

Because Eq.~\eqref{eq:direct-space-charge-gaussian} requires integration around the entire circumference, evaluating the overall direct tune shift requires detailed information of the ring lattice functions.

Fig.~\ref{figs:HSR_SpaceCharge} depicts the direct space-charge incoherent tune footprint for the $100\text{ GeV}$ Hadron Storage Ring (HSR) proton lattice, operating with a bunch intensity of $6.9 \times 10^{10}$ protons and an RMS bunch length of $\sigma_z = 70\text{ mm}$. 

This yields direct space-charge incoherent tune shifts $\Delta \nu_{x, \text{incoh}}^{\text{sc}} = -0.008$ and $\Delta \nu_{y, \text{incoh}}^{\text{sc}} = -0.025$. These numerical results show reasonable agreement with the analytical estimates of $\Delta \nu_{x, \text{incoh}}^{\text{sc}} = -0.004$ and $\Delta \nu_{y, \text{incoh}}^{\text{sc}} = -0.021$ calculated using Eq.~\eqref{eq:direct-space-charge-gaussian}. The factor of 2 discrepancy in the horizontal space charge tune shift likely comes from lattice-specific effects ($\beta$-function variation and horizontal dispersion) rather than numerical inaccuracies in the simulation.

In addition to the direct space-charge incoherent tune shift formula for a bunched bi-Gaussian elliptical beam, alternative expressions for uniform and Gaussian distributions across various beam geometries are widely documented in the literature~\cite{Schindl2006_SpaceCharge}. These expressions are summarized in Appendix~\ref{sec:direct-space-charge-formula} for completeness. These expressions provide useful benchmark estimates during the initial collider design phase, when detailed lattice parameters are not yet available.

\subsection{\label{sec:indirect-space-charge} Indirect Space-Charge-Induced Quadrupolar Incoherent Tune Shifts}

This section examines the indirect space-charge incoherent tune shift arising from non-circular vacuum chamber boundaries around the ring.

For a rotationally symmetric circular beam pipe, boundary-induced image forces produce no net indirect incoherent space-charge tune shift. In contrast, the vacuum chamber in the HSR and ESR has a rectangular cross-sectional profile. The resulting broken rotational symmetry induces quadrupolar components of the image force and incoherent tune shifts.

Two primary approaches exist to evaluate the indirect space-charge incoherent tune shift from the surrounding vacuum chamber. One is the classical electric and magnetic image-charge method originally developed by Laslett ~\cite{Laslett1963, Ng2001_TM2152, Ng2001_TM21367, Ng2005_Book, Zotter1975_QShift}.  In addition to the Laslett formulation, the indirect space-charge incoherent tune shift can also be evaluated from the transverse wake function or transverse impedance of a resistive-wall vacuum chamber, using either analytical models (e.g., A. Chao~\cite{ChaoHeifetsZotter2002}, A. Blednykh~\cite{Blednykh2016_LowFrequencyQuadrupoleImpedance}, 
Y. Shobuda ~\cite{ShobudaYokoya2002_ResistiveWall}) or numerical simulation codes~\cite{Gunzel2004_WEPLT083, Gunzel2006_PRSTAB_114402}.

The theory of wakefield-induced incoherent tune shifts was briefly discussed in Refs.~\cite{Brunelle2016_PRAB_044401, Wang2022Quadrupolar}. Here, we extend these formulations and recast them in a consistent form for rectangular chamber geometries, facilitating a direct comparison among the different approaches considered in this paper.

\begin{figure}[!ht]
\centering
\includegraphics[width=\textwidth]{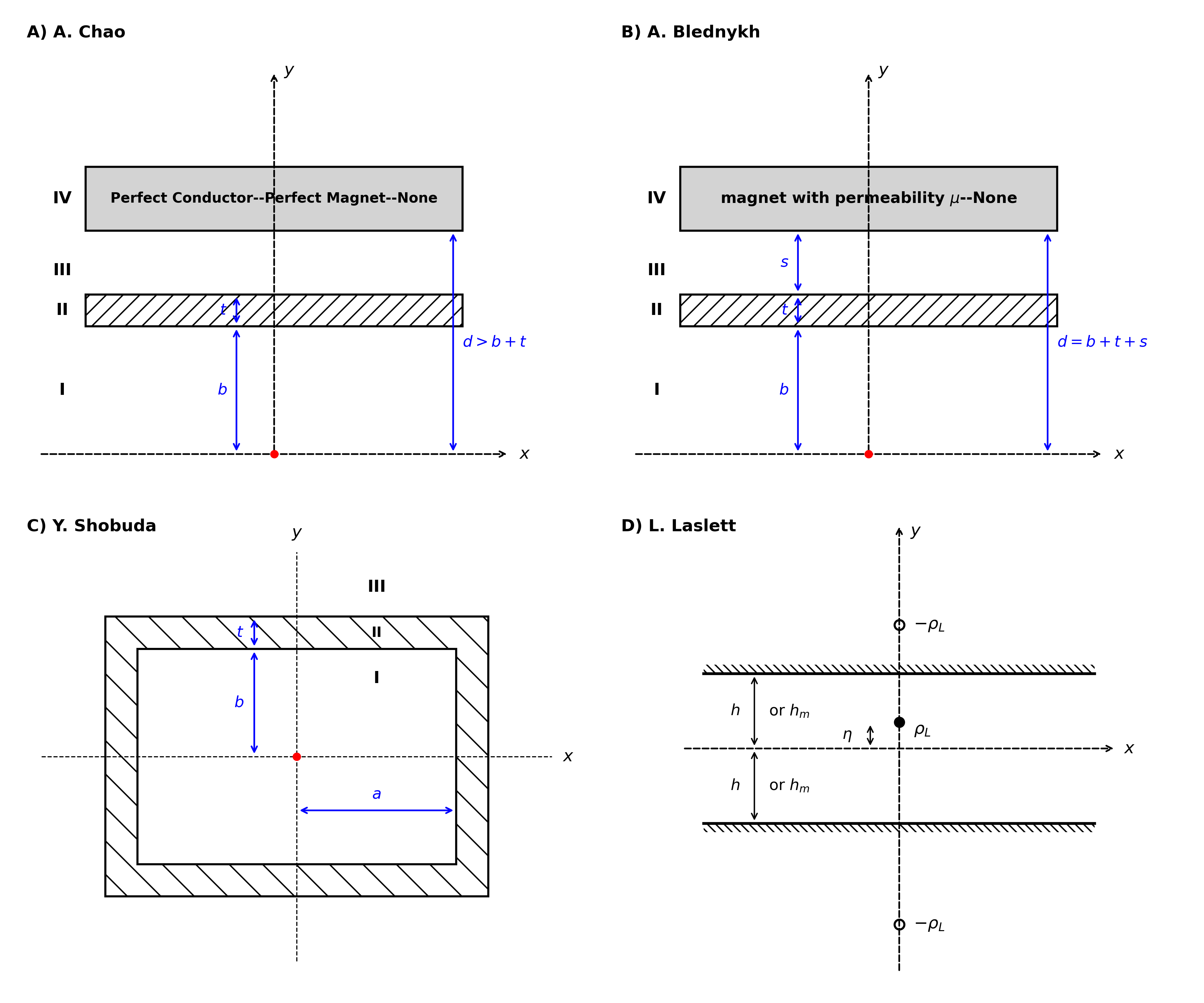}
\caption{Boundary conditions and geometry representations across four analytical models for quadrupolar wakefield and impedance calculations: (Top Left) A. Chao model using parallel plates with idealized perfect conductor/magnet boundaries; (Top Right) A. Blednykh model incorporating a finite-thickness wall and magnet poles with finite permeability $\mu$; (Bottom Left) Y. Shobuda model for a 2D rectangular vacuum chamber with finite wall thickness $t$; (Bottom Right) L. Laslett classical model using parallel planes and the method of image charges/currents.}
\label{figs:Image_SC_4Method}
\end{figure}

Fig.~\ref{figs:Image_SC_4Method} shows the boundary-condition configurations for the quadrupolar impedance incoherent tune-shift models considered in four analytical approaches. The model by A. Chao et al. assumes a parallel-plate chamber with half-height $b$ and wall thickness $t$, with the outer boundary at $d>b+t$ represented by either a perfect electric conductor, a perfect magnetic conductor, or no outer wall ($d\rightarrow\infty$). The model by A. Blednykh et al. considers magnet poles located at $d=b+t+s$ and explicitly accounts for the finite permeability $\mu$ of the magnet poles, or no outer wall ($d\rightarrow\infty$). The model by Y. Shobuda and K. Yokoya considers a fully enclosed two-dimensional elliptical or rectangular chamber with half-dimensions $a\times b$ and finite wall thickness $t$, accounting for low-frequency field penetration into the outer vacuum region. Finally, the classical model by L. Laslett employs a parallel-plate configuration with half-gap $h$ (or magnetic-pole half-height $h_m$) and evaluates the field using an infinite array of image charge densities $-\rho_L$ for a line source displaced vertically by $\eta$. The Laslett formulation can be applied to parallel-plate, rectangular, and elliptical chamber geometries.

\subsubsection{Chao Formalism}

Under the assumption that the long-range resistive-wall wake dominates the tune shift, A. Chao formulated the indirect space-charge incoherent tune shift as a quadrupolar impedance effect. By summing the wakefield contributions across all preceding bunches in a train under the specific outer-boundary conditions illustrated in Fig.~\ref{figs:Image_SC_4Method}, the quadrupolar resistive-wall incoherent tune shift for a noncircular chamber developed by A. Chao is expressed as~\cite{ChaoHeifetsZotter2002}:

\begin{equation}
\begin{aligned}
\label{eq:Chao_r_cases}
\frac{\Delta \nu_{x,y}}{dI} &= \pm \left( \frac{\pi }{48 \nu_{x,y}} \right) \left( \frac{Z_0}{E/e} \right) \left( \frac{R}{b} \right)^2 \left( \frac{L}{C} \right)r \\
r &=
\begin{cases}
1+\left(\dfrac{b}{d}\right)^2,
& d> b+t \quad\text{perfect magnet}, \\[1ex]
1-\left(\dfrac{b}{d}\right)^2,
& d> b+t \quad\text{perfect conductor}, \\[1ex]
1,
& d\rightarrow\infty \quad \text{no outer wall}.
\end{cases}
\end{aligned}
\end{equation}

where $r$ is the low-frequency shielding factor, $b$ is the half-height of the rectangular vacuum chamber, $d$ is the distance from the beam axis to the outer boundary (magnet pole or conductor), and $t$ is the thickness of the vacuum-chamber wall. In this expression, $\nu_{x,y}$ denotes the horizontal or vertical betatron tune, $I$ is the average beam current, $Z_0 = 120\pi\,\Omega $ is the impedance of free space, and $E$ is the total beam energy, so that $E/e$ is the beam energy expressed in volts, where $e$ is the elementary charge.

The parameter \(R\) represents the average machine radius. In general, \(L\) denotes the length of the rectangular resistive-wall section, which may have either a dipole magnet or a vacuum chamber without any additional outer wall as its boundary condition. The ratio \(L/C\) represents the fraction of the total ring circumference occupied by the non-circular vacuum-chamber section with the specified boundary condition.

The resulting tune shift is independent of the vacuum chamber wall thickness $t$ and material conductivity $\sigma_c$. The resistivity and wall thickness determine how quickly the system reaches that limit, but not the final static focusing coefficient. The final long-term incoherent tune shift is actually the Laslett-type static image-field result. Over sufficiently long times, the accumulated wake approaches the static image-charge/image-current field, which is why the result becomes independent of resistivity and wall thickness.

According to Eq.~\eqref{eq:Chao_r_cases} and Fig.~\ref{figs:Image_SC_4Method}, the Chao formalism can be applied to beam pipes with a perfectly conducting outer boundary. This assumption is well suited to the HSR beam-pipe configuration described above, in which the existing stainless-steel RHIC beam pipe serves as the outer structural boundary surrounding the inserted rectangular beam screen.

\subsubsection{Blednykh Formalism}

Starting from the complex betatron frequency shift of a coupled-bunch mode using a set of $M$ (bunch number) coupled Vlasov equations and the general expression for the quadrupole impedance at zero frequency—under the outer-boundary conditions illustrated in Fig.~\ref{figs:Image_SC_4Method}—A. Blednykh developed a formula for the quadrupole incoherent tune shift with an outer magnetic boundary~\cite{Blednykh2016_LowFrequencyQuadrupoleImpedance}:
\begin{equation}
\label{eq:quadrupole_impedance}
\begin{aligned}
Z_{Q_{x,y}}(0) &= \pm i \frac{\pi^2}{12 c b^2} \left(1 + 2 \frac{b^2}{d^2} f(\eta)\right)\\
f(\eta) &= \frac{6}{\pi^2} \mathrm{Li}_2(\eta)\\
\eta &= \frac{\mu_r - 1}{\mu_r + 1}.
\end{aligned}
\end{equation}
where $c$ is the speed of light, $b$ has the same definition as above, and $d = b+t+s$ is explicitly the distance from the beam axis to the inner surface of the magnet, i.e., the half-magnetic gap. Blednykh introduces a continuous material-dependent boundary factor $f(\eta)$ to replace a binary boundary assumption with a finite-permeability interpolation. Thus, whereas the Chao model represents the outer boundary using discrete cases, the Blednykh formulation incorporates the magnetic permeability through the continuous function $f(\eta)$. 

In particular, the incoherent tune shift is formulated in terms of the zero-frequency quadrupolar impedance of the vacuum chamber with magnetic boundaries. For a localized chamber of length $L$, the horizontal or vertical tune shift induced by the quadrupolar impedance is given by~\cite{Blednykh2016_LowFrequencyQuadrupoleImpedance}:
\begin{equation}
\Delta \nu_{x,y} = \frac{I_{\rm av} L}{4 \pi E / e} \, \beta_{x,y} \, \mathrm{Im} \, Z_{Q_{x,y}}(0)
\end{equation}

After converting the impedance per unit length from CGS units to SI units by multiplying by $cZ_0/(4\pi)$, and replacing $\beta_{x,y}=R/\nu_{x,y}$, the expression becomes:
\begin{equation}
\label{eq:Alexei-tune-shifts-incoherent}
\begin{aligned}
\frac{\Delta \nu_{x,y}}{dI} &= \pm \left( \frac{\pi }{48 \nu_{x,y}} \right) \left( \frac{Z_0}{E/e} \right) \left( \frac{R}{b} \right)^2 \left( \frac{L}{C} \right) r \\
r &= \begin{cases}
 \frac{1}{2}+\frac{b^2}{d^2}f(\eta) & \text{ magnet }  d = b+t+s\\[1ex]
\frac{1}{2}  & \text{no magnet } d \rightarrow\infty \quad \\[1ex]
\end{cases}\\
f(\eta) &= 
\begin{cases}
1 , & \text{perfect magnet } \quad \mu\rightarrow \infty \\[1ex]
0.99, & \text{soft magnet}  \quad \mu = 400\\[1ex]
0, & \text{saturated magnet} \quad \mu = 1
\end{cases}
\end{aligned}
\end{equation}

This expression has the same overall dependence on beam energy, chamber aperture, lattice radius, and fractional chamber length as the Chao expression, but the geometrical/material factor $r$ is different. For the outside of the dipoles, there are no magnet poles ( $d\rightarrow\infty$, $f = 0$), so $r=\frac{1}{2}$ applies to the remaining section of the ring.

Here, the smooth approximation for the beta function is used to obtain Eq.~\eqref{eq:Alexei-tune-shifts-incoherent} through the following relation, allowing direct comparison with the formulation by A. Chao:
\begin{equation}
\langle\beta_{x,y}\rangle
\simeq
\frac{R}{\nu_{x,y}},
\end{equation}

When the lattice is available, the actual beta function integrated around the ring should be used. 
\begin{equation}
\langle\beta_{x,y}\rangle_L=
\frac{1}{L}
\int_L\beta_{x,y}(s),ds .
\end{equation}

\subsubsection{Shobuda Formalism}

By solving Maxwell's equations for a wall of finite thickness $t$, Y. Shobuda et al~\cite{ShobudaYokoya2002_ResistiveWall} derived the indirect space-charge incoherent tune shift arising from the resistive-wall wake function of vacuum chambers with finite wall thickness for both elliptical and rectangular cross sections. For a rectangular vacuum chamber, the tune shift can be expressed as follows~\cite{ShobudaYokoya2002_ResistiveWall}:
\begin{equation}
\begin{aligned}
\label{eq:Shobuda}
\frac{\Delta \nu_{\text{inc}, x,y}}{dI} &\approx \pm D_{2xy} \cdot \frac{L \langle \beta_{x,y} \rangle}{4\pi} \cdot \frac{Z_0}{E} \cdot \frac{1}{\pi b^3} \cdot t \\
&\approx \pm \left(\frac{1}{2\pi\nu_{x,y}} \right)
\left( \frac{Z_0}{E/e} \right)
\left( \frac{R}{b} \right)^2
\left( \frac{L}{C} \right)
\cdot \frac{t}{b} \cdot D_{2xy}\\
D_{2xy} &= \frac{\pi^3}{8} \sum_{\substack{m=1 \\ m \text{ odd}}}^{\infty} \left( \frac{m^2}{\cosh^2 \left( \frac{m \pi a}{2b} \right)} - \frac{b^3 m^2}{a^3 \cosh^2 \left( \frac{m \pi b}{2a} \right)} \right)
\end{aligned}
\end{equation}
where $t$ is the thickness of the vacuum chamber wall, $a$ and $b$ are the inner half-width and half-height of the rectangular chamber. 

From Eq.~\eqref{eq:Shobuda}, we can see that the indirect space-charge incoherent tune shift depends on the vacuum-chamber wall thickness $t$ and the width $a$ of the rectangular chamber, whereas the formulas derived by A. Chao and A. Blednykh do not explicitly include these parameters.

Eq.~\eqref{eq:Shobuda} also neglects the effects of surrounding materials—particularly magnetic materials—because it assumes that fields leaking out of the chamber can be neglected on sufficiently long time scales. Furthermore, this expression assumes that the beam-pipe dimensions are sufficiently larger than the skin depth $\delta_{\mathrm{skin}}$ and that $k\rho \ll 1$, where $k$ is the free-space wave number and $\rho$ is the normalized resistivity.

\subsubsection{Laslett Formalism}

After reviewing the approaches described above, the classical formulation developed by Laslett is adopted in this work to evaluate the incoherent tune shifts in the HSR and ESR for two primary reasons. First, although some experimental comparisons have been reported for other formulations~\cite{ChaoHeifetsZotter2002,Sakanaka2005_PRSTAB_042801,Migliorati2018_PRAB_041001}, the Laslett model provides a conservative, upper-bound estimate of the incoherent tune shift~\cite{Brunelle2016_PRAB_044401}. Second, the Laslett approach explicitly accounts for off-center beam trajectories, which is particularly important for the HSR, where the beam is expected to operate with an offset of approximately $20\text{ mm}$ from the chamber center for 100 GeV ion beams. At 5 GeV, the ESR beam passes through the reverse-bend dipoles with a horizontal orbit offset of about $±10\text{ mm}$.

As shown in Fig.~\ref{figs:Image_SC_4Method}, the Laslett calculation uses an idealized image-charge and image-current boundary condition at the specified beam aperture. It does not include an additional outer-conductor boundary or magnet poles.

The original Laslett expressions of indirect space-charge incoherent tune shifts for rectangular beam chambers are presented below~\cite{Ng2001_TM2152}:

\begin{equation}
\begin{aligned}
\Delta \nu_{\text{(y,x), incoh}}^{sc, indirect}
={}&-\frac{N r_0 R}
{\pi\gamma\beta^2\nu_0^{y,x}}
\Bigg[
\left(
\frac{1-\chi_e-\beta^2}{B_f}+\beta^2
\right)
\frac{\epsilon_1^{y,x}}{h^2}
\\
&+
\mathcal{F}_{dipole}\beta^2
\frac{\epsilon_2^{y,x}}{h_m^2}
+
(1-\chi_e-\beta^2)
\frac{\epsilon_{\text{spch}}^{y,x}}
{a_{y}^2}
\Bigg]
\\
={}&-\frac{N_{\text{total}}r_0R}
{\pi\gamma\beta^2\nu_0^{y,x}}
\left[
\frac{\epsilon_1^{y,x}}{h^2}
+
\mathcal{F}_{dipole}
\frac{\epsilon_2^{y,x}}{h_m^2}
\right].
\end{aligned}
\label{eq:indirect-space-charge-incoherent}
\end{equation}
where, $N$ is the total number of particles in the whole beam; $B_f$ is the bunching factor or the ratio of average to peak current; $\chi_e$ is neutralization factor; $\nu_0^{y,x}$ is lattice bare betatron tune, $h$ is half vertical height of the vacuum chamber, $h_m$ is the half-gap between magnet poles. 

Here, $\mathcal{F}_{dipole}$ is the filling factor of the vacuum chamber enclosed by magnet poles or dipole magnets.  With the latest lattice,  $\mathcal{F}_{dipole} = 0.417$ for HSR and $\mathcal{F}_{dipole} = 0.442$ for ESR. 

Furthermore, $\epsilon_1^{x,y}$ and $\epsilon_2^{x,y}$ represent the incoherent electric and magnetic Laslett image coefficients, respectively. For a rectangular vacuum chamber enclosed by saturated superconducting magnet poles (where magnetic image effects vanish, i.e., $\epsilon_2^y = \epsilon_2^x = 0$), the electric Laslett image coefficients are given by~\cite{Ng2001_TM2152}:
\begin{equation}
\label{eq:laslett-tune-shifts-incoherent_orignal}
\begin{aligned}\epsilon_1^y &= -\epsilon_1^x = \frac{K^2(k)}{12} \left(1 - 6k + k^2\right)
\end{aligned}
\end{equation}
\noindent
\noindent where $w/h = K(k')/(2K(k))$ for the rectangular cross section, with $w$ and $h$ representing the half-width and half-height of the vacuum chamber, $k' = \sqrt{1 - k^2}$, and $K' = K(k')$. Therefore, the electric image field is vertically defocusing but horizontally focusing (with the sign of the image term changing), which is a typical feature of rectangular or elliptical vacuum pipes.

The Laslett incoherent tune shift decreases with $\frac{1}{\gamma}$, much slower than the direct space-charge term $\frac{1}{\gamma^3}$, and thus it is therefore a concern for electron and high-energy proton machines.


For comparison with Eqs.~\eqref{eq:Chao_r_cases},~\eqref{eq:Alexei-tune-shifts-incoherent}, and~\eqref{eq:Shobuda}, Eq.~\eqref{eq:indirect-space-charge-incoherent} can be further modified by accounting for the fraction of the ring occupied by the rectangular beam pipe, yielding
\begin{equation}
\label{eq:laslett-tune-shifts-incoherent_modified}
\begin{aligned}
\frac{\Delta \nu_{x,y}}{dI}
&=
\pm
\left( \frac{1}{2\pi \beta^3 \nu_{x,y}} \right)
\left( \frac{Z_0}{E/e} \right)
\left( \frac{R}{h} \right)^2 r\\
r &= 
\mathcal{F}_{rect}\epsilon_{1}^{x,y}
+
\mathcal{F}_{dipole}\,\epsilon_{2}^{x,y}
\frac{h^2}{h_m^2}
.
\end{aligned}
\end{equation}

Here, \(\mathcal{F}_{\mathrm{rect}}\) denotes the fraction of the ring circumference occupied by the rectangular beam pipe. We take \(\mathcal{F}_{\mathrm{rect}}=0.79\) for the HSR and \(\mathcal{F}_{\mathrm{rect}}=1\) for the ESR, since only a small fraction of the ESR has a round beam pipe.

When $N$ is converted to the total circulating beam current $I$, an additional factor of $1/\beta$ is introduced through $N = \frac{2\pi RI}{e\beta c}$. Therefore, the $1/\beta^2$ dependence becomes $1/\beta^3$ in the current-based Eq.~\eqref{eq:laslett-tune-shifts-incoherent_modified}.

\begin{figure}[!ht]
\centering
\includegraphics[width=0.8\columnwidth]{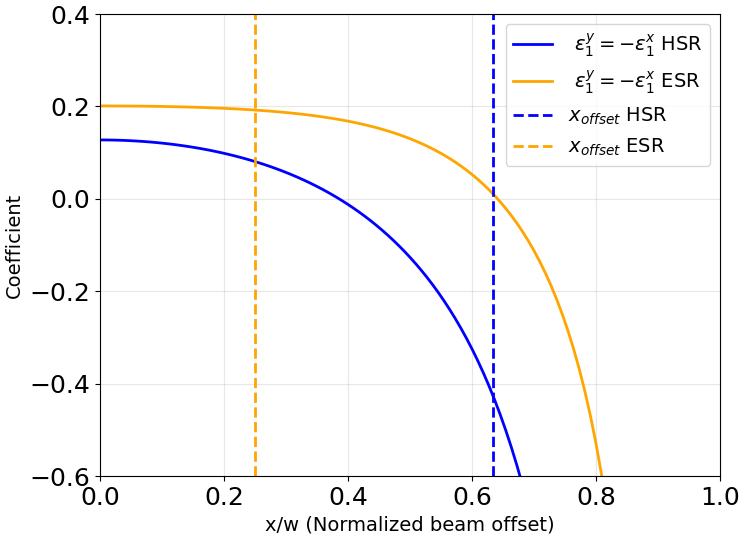}
\caption{Laslett coefficient $\varepsilon_{1}^{y}=-\varepsilon_{1}^{x}$ for a rectangular vacuum chamber with an offset beam: (a) HSR with $h/w=24/31.55$ and a $20~\mathrm{mm}$ offset ($x/w=20/31.55$), yielding $\varepsilon_{1}^{y}=-\varepsilon_{1}^{x}=-0.427$; and (b) ESR with $h/w=18/40=0.45$ and a $10~\mathrm{mm}$ offset ($x/w=10/40$), yielding $\varepsilon_{1}^{y}=-\varepsilon_{1}^{x}=0.192$, which is close to the centered-beam value of $0.201$. Therefore, the ESR Laslett tune shifts with beam offset are not calculated separately. The above $\varepsilon_{1}^{y}=-\varepsilon_{1}^{x}$ values are used for evaluating the HSR incoherent Laslett tune shifts in Table~\ref{tb:laslett-tune-shifts-incoherent}.}
\label{figs:Laslett_Coeffiecient}
\centering
\end{figure}

\begin{table}[htbp]
\centering
\footnotesize
\begin{tabular}{c|l|c|c|c}
\hline
\textbf{} & \textbf{Parameter} & \textbf{5 GeV} & \textbf{10 GeV} & \textbf{18 GeV} \\
\hline
Ring circumference & $C$ & \multicolumn{3}{c}{3833~$\mathrm{m}$} \\
Average ring radius & $R=C/(2\pi)$ & \multicolumn{3}{c}{609.77~$\mathrm{m}$} \\
Half-Height & $b$ or $h$ [m] & \multicolumn{3}{c}{0.018} \\
Half-Width & $w$ [m] & \multicolumn{3}{c}{0.040} \\
Half-height of Magnet Pole & $h_m$ [m] & \multicolumn{3}{c}{0.026} \\
Thickness & $t$ [m] & \multicolumn{3}{c}{0.004} \\
& $\varepsilon_{1}^y=-\varepsilon_{1}^x$ & \multicolumn{3}{c}{0.201} \\
& $\varepsilon_{2}^y=-\varepsilon_{2}^x$ & \multicolumn{3}{c}{0.411} \\
& $\mathcal{F}_{\mathrm{Dipole}}$ & \multicolumn{3}{c}{0.442} \\
& $\mathcal{F}_{\mathrm{Rect}}$ & \multicolumn{3}{c}{1} \\
& $I$ [A] & \multicolumn{3}{c}{1} \\
Lattice Bare Horizontal Tune & $Q_x$ & 39.08 & 38.08 & 45.08 \\
Lattice Bare Vertical Tune & $Q_y$ & 34.14 & 33.14 & 41.14 \\
\hline
\hline
Laslett & $d\nu_x/dI $ & 0.1016 & 0.0522 & 0.0245 \\
& $d\nu_y/dI $ & -0.1163 & -0.0599 & -0.0268 \\
\hline
\hline
\end{tabular}
\caption{Comparison of the EIC ESR parameters at 5 GeV, 10 GeV, and 18 GeV for 1 A current. All machine parameters are identical except for the beam energy. The Laslett coefficients $\varepsilon_{1}^{y}=-\varepsilon_{1}^{x}$ were calculated for a rectangular vacuum chamber, while the coefficients $\varepsilon_{2}^{y}=-\varepsilon_{2}^{x}$ were calculated for the parallel-plate approximation. In the table, the normalized tune slopes $\mathrm{d}\nu_x/\mathrm{d}I$ and $\mathrm{d}\nu_y/\mathrm{d}I$ are listed in units of $\text{A}^{-1}$.}
\label{tab:eic_esr_5_10_18gev}
\end{table}

\begin{table}[htbp]
    \centering
    \begin{tabular}{l|l|c|c|c}
        \toprule
        \textbf{Model} & \textbf{Image} & $\mathbf{\frac{L}{C}}$ & $\mathrm{d}\nu_x/\mathrm{d}I$ ($\mathrm{A}^{-1}$) & $\mathrm{d}\nu_y/\mathrm{d}I$ ($\mathrm{A}^{-1}$) \\
        \midrule
        A. Chao        & Open              & 0.558 & 0.0809 & -0.0926 \\
        A. Chao        & Magnetic          & 0.442 & 0.0948 & -0.1086 \\
        A. Chao        & Magnetic + Open   & 1     & 0.1757 & -0.2012 \\        
        A. Blednykh    & Open              & 0.558 & 0.0405 & -0.0463 \\        
        A. Blednykh    & Magnetic          & 0.442 & 0.0628 & -0.0719 \\
        A. Blednykh    & Magnetic + Open   & 1     & 0.1033 & -0.1182 \\        
        Y. Shobuda     & Electric          & 1     & 0.0329 & -0.0377 \\
        L. Laslett     & Electric+magnetic & 1     & 0.1016 & -0.1163 \\
        \bottomrule
    \end{tabular}
    \caption{Comparison of indirect space-charge tune slopes for the ESR 5 GeV lattice using different analytical models and chamber-boundary assumptions. The chamber dimensions are $h = 18\text{ mm}$ and $w = 40\text{ mm}$, where $d = h_m = 26\text{ mm}$ represents the distance to the magnet pole, see Table \ref{tab:eic_esr_5_10_18gev}. In the table, the normalized tune slopes $\mathrm{d}\nu_x/\mathrm{d}I$ and $\mathrm{d}\nu_y/\mathrm{d}I$ are listed in units of $\text{A}^{-1}$.} 
\label{tab:tune_comparison_5GeV_ESR}
\end{table}

\begin{table}[htbp]
\centering
\begin{tabular}{l|l|c|c|c|c}
\hline\hline
Energy & Parameter &
Rect 1 &
Rect 2 &
Elliptical &
Rect 2($20$mm Offset) \\
\hline

& $h$ (m) & 0.0240 & 0.0240 & 0.0256 & 0.0240 \\

& $w$ (m) & 0.0280 & 0.03155 & 0.0336 & 0.03155 \\

& $h_{\mathrm{m}}$ (m) &
\multicolumn{4}{c}{0.040} \\

& $\varepsilon_{1}^{y}=-\varepsilon_{1}^{x}$ &
0.08219 & 0.12744 & 0.10306 & -0.42700 \\

& $\varepsilon_{2}^{y}=-\varepsilon_{2}^{x}$ &
\multicolumn{4}{c}{0} \\

& $\mathcal{F}_{\mathrm{Dipole}}$ &
\multicolumn{4}{c}{0.417} \\

& $\mathcal{F}_{\mathrm{Rect}}$ &
\multicolumn{4}{c}{0.79} \\

& $I$ [A] &
\multicolumn{4}{c}{1} \\

 & $Q_x$ &
\multicolumn{4}{c}{27.228} \\
 & $Q_y$ &  
\multicolumn{4}{c}{25.210} \\

\hline

23.8 GeV & $d\nu_x/dI $ & 0.00389 & 0.00603 & 0.00428 & -- \\
& $d\nu_y/dI$ & -0.00420 & -0.00651 & -0.00463 & -- \\

\hline

41 GeV & $d\nu_x/dI $ & 0.00226 & 0.00350 & 0.00249 & -- \\
& $d\nu_y/dI$ & -0.00244 & -0.00378 & -0.00268 & -- \\

\hline

100 GeV & $d\nu_x/dI $ & 0.00064 & 0.00143 & 0.00070 & -0.00480 \\
& $d\nu_y/dI$ & -0.00069 & -0.00155 & -0.00076 & 0.00519 \\

\hline

275 GeV & $d\nu_x/dI $ & 0.0003 & 0.0005 & 0.0004 & -- \\
& $d\nu_y/dI $ & -0.0004 & -0.0006 & -0.0004 & -- \\

\hline\hline
\end{tabular}
\caption{Incoherent Laslett tune shifts (normalized to 1 A) for HSR, comparing rectangular vacuum chambers with half-widths of $28.00$ and $31.55$ mm, a rectangular chamber with a $20$ mm beam offset, and an elliptical vacuum chamber ($h=25.6$ mm and $w=33.6$ mm). In the table, the normalized tune slopes $\mathrm{d}\nu_x/\mathrm{d}I$ and $\mathrm{d}\nu_y/\mathrm{d}I$ are listed in units of $\text{A}^{-1}$.}
\label{tb:laslett-tune-shifts-incoherent}
\end{table}

For the ESR, the indirect space-charge incoherent tune shifts evaluated via the Laslett formalism at 5, 10 and 18~GeV for a beam current of 1~A are listed in Table~\ref{tab:eic_esr_5_10_18gev}. 

Table~\ref{tab:tune_comparison_5GeV_ESR} compares the calculated horizontal and vertical tune slopes, $d\nu_x/dI$ and $d\nu_y/dI$, for the ESR 5~GeV lattice using different analytical models. For the A.~Chao and A.~Blednykh models, the perfectly conducting magnetic-boundary condition is applied over the dipole regions, using the ESR dipole filling factor \(\mathcal{F}_{\mathrm{dipole}}=0.442\). For comparison, the A.~Chao and A.~Blednykh models are also evaluated with an open outer boundary (no outer wall, $ d \rightarrow\infty$), with \(r=1\) and $\frac{L}{C} = 0.558$ as well as the full-ring result, which combines both contributions. In contrast, the Y.~Shobuda model does not explicitly include a dipole filling factor and is therefore evaluated over the full ring circumference, \(\frac{L}{C}=1\). The L.~Laslett model includes both the electric-image contribution from the full ring and the magnetic-image contribution from the dipoles.  

Table~\ref{tab:tune_comparison_5GeV_ESR} shows that A. Blednykh's full-ring model yields $0.1033\text{ A}^{-1}$ and $-0.1182\text{ A}^{-1}$, which agree with Laslett's values ($0.1016\text{ A}^{-1}$ and $-0.1163\text{ A}^{-1}$) to within $2\%$. 

While  A. Chao (full ring) exceeds them because the A. Chao open-boundary term ($r = 1$) corresponds to an image coefficient of \(\pi^2/24\), twice the Laslett parallel-plate electric coefficient of \(\pi^2/48\). With \(r = 1/2\), as in Blednykh’s formalism, the Chao and Laslett results agree. A similar factor-of-two discrepancy has been noted in Refs~\cite{Blednykh2016_LowFrequencyQuadrupoleImpedance}. Nevertheless, all models give a positive horizontal tune slope, $d\nu_x/dI$, and a negative vertical tune slope, $d\nu_y/dI$.  

For the off-center beam, the Laslett coefficients $\varepsilon_{1}^{xy}$ for a rectangular chamber are calculated using the formula in Ref.~\cite{Ng2005_Book} and plotted in Fig.~\ref{figs:Laslett_Coeffiecient} for the HSR beam screen and ESR vacuum chamber. Additional plots of Laslett coefficients for off-center beam are presented in Appendix~\ref{sec:Off-center-Laslett}. 

In this paper, the transverse boundary of the HSR beam screen can be defined by a rectangular cross-section (Fig.~\ref{figs:HSR_Screen}) with a vertical half-height of $a = 24.00\,\mathrm{mm}$ and a horizontal half-width of $b = 31.55\,\mathrm{mm}$, corresponding to a maximum outer diameter of $\varnothing 63.10\,\mathrm{mm}$. Alternatively, the rectangular boundary can be defined with $b = 28\,\mathrm{mm}$ while maintaining the same cross-sectional area. An elliptical geometry can also be considered, with a vertical half-height of $h = 25.6\,\mathrm{mm}$ and a horizontal half-width of $w = 33.6\,\mathrm{mm}$. The three transverse-boundary models are compared below in terms of their impact on the incoherent tune shift. Based on the comparison, the first rectangular geometry, with $a = 24.00\,\mathrm{mm}$ and $b = 31.55\,\mathrm{mm}$, is adopted for the subsequent HSR evaluation. The complex geometry of the ESR vacuum chamber (Fig.~\ref{figs:HSR_Screen}) is approximated as an equivalent rectangular chamber with a half-height of $a = 18.00\text{ mm}$, a half-width of $b = 40\text{ mm}$, and a wall thickness of $t = 4\text{ mm}$.

The indirect space-charge incoherent tune shifts per unit current ($1~\mathrm{A}$) for HSR proton beams—evaluated across different rectangular definitions, both with and without orbit offsets—are summarized in Table~\ref{tb:laslett-tune-shifts-incoherent}. From the table, the rectangular model with a half-width of \(31.55\text{ mm}\) exhibits larger tune shifts than both the elliptical model and the rectangular model with a \(28\text{ mm}\) half-width. Therefore, the \(31.55\text{ mm}\) rectangular configuration will be used in the remainder of this paper.

The relevant beam parameters, including the incoherent tune shifts calculated using the corresponding beam currents and arising from beam--beam interactions and space-charge effects, are summarized in Table~\ref{tab:eic_unified_summary}. These results provide a basis for assessing the required amplitude-dependent tune spread and compensation capability by adjusting the machine sextupoles and octupoles~\cite{Drees2024_BetatronDetuning}. The Laslett horizontal tune shift calculated for the ESR in this table agrees well with the value reported in the PDR~\cite{Peggs2026_EICPDR}: $\Delta Q_x = 0.254$ in this work, compared with $\Delta Q_x = 0.242$ in the PDR. The vertical tune shift calculated in this work for an ESR beam current of $2.5\,\mathrm{A}$ is $\Delta Q_y = -0.291$; the corresponding value is not reported in the PDR. The electron beam-beam parameter (0.1765) for the 100 GeV–5 GeV vertical collision configuration still will be performed with updated parameters.

\begin{table}[htbp]
\centering
\resizebox{\textwidth}{!}{%
\begin{tabular}{l|c|c|c|c|c|c|c}
\hline
\textbf{Parameter} & \textbf{Unit} & \textbf{Injection} & \textbf{41-5 GeV} &
\textbf{100-5 GeV} & \textbf{100-10 GeV} & \textbf{275-10 GeV} &
\textbf{275-18 GeV} \\
\hline
Proton Energy $E_p$   & GeV & 23.8 & 41  & 100 & 100 & 275 & 275 \\
Electron Energy $E_e$ & GeV & - & 5   & 5   & 10  & 10  & 18  \\
No. of bunches $M$ & -- & 290 & 1160 & 1160 & 1160 & 1160 & 290 \\
Required beam size $\sigma_x^*$ & $\mu$m & - & 198.00 & 125.00 & 138.00 & 95.00 & 119.00 \\
Required beam size $\sigma_y^*$ & $\mu$m & - & 27 & 11.00 & 12.00 & 8.50 & 11.00 \\
\hline
\multicolumn{8}{l}{\textbf{Protons HSR}} \\
\hline
Bunch intensity & $10^{10}$ & 27.6 & 2.6 & 4.8 & 6.9 & 6.9 & 19.1 \\
Beam current & A & 1.0 & 0.38 & 0.69 & 1.0 & 1.00 & 0.69 \\
RMS normalized emittance H & $\mu$m & 3.0 & 1.9 & 2.7 & 3.2 & 3.3 & 5.2 \\
RMS normalized emittance V & $\mu$m & 0.3 & 0.45 & 0.25 & 0.29 & 0.30 & 0.47 \\
RMS emittance H & nm & 118.3 & 44 & 26 & 30.0 & 11.3 & 18.0 \\
RMS emittance V & nm & 11.8 & 10 & 2.3 & 2.7 & 1.0 & 1.6 \\
RMS bunch length $\sigma_l^*$ & mm & 1500 & 75 & 70 & 70 & 60 & 60 \\
RMS $\delta_p/p$ & $10^{-3}$ & 1.0 & 1.03 & 0.97 & 0.97 & 0.68 & 0.68 \\
Collision $\beta^*$ H & m & - & 0.90 & 0.61 & 0.63 & 0.80 & 0.80 \\
Collision $\beta^*$ V & m & - & 0.071 & 0.055 & 0.057 & 0.072 & 0.071 \\
Beam-beam parameter $\xi_H$ & -- & - & 0.0150 & 0.0138 & 0.0187 & 0.00958 & 0.0023 \\
Beam-beam parameter $\xi_V$ & -- & - & 0.0087 & 0.0106 & 0.0195 & 0.00984 & 0.0025 \\
Space charge tune shift $\Delta \nu_x$ & -- & -0.020 & -0.015 & -0.003 & -0.004 & -0.001 & -0.001 \\
Space charge tune shift $\Delta \nu_y$ & -- & -0.082 & -0.043 & -0.016 & -0.021 & -0.003 & -0.006 \\
Laslett tune shift $\Delta \nu_x$ & -- & 0.00603 & 0.00133 & 0.00099 & 0.00143 & 0.0005 & 0.0004 \\
Laslett tune shift $\Delta \nu_y$ & -- & -0.00651 & -0.00144 & -0.00107 & -0.00155 & -0.0006 & -0.0004 \\
Proton beam size $\sigma_x^*$ & $\mu$m & - & 189.74 & 125.94 & 137.48 & 95.08 & 120.00 \\
Proton beam size $\sigma_y^*$ & $\mu$m & - & 26.65 & 11.25 & 12.41 & 8.49 & 10.66 \\
Lattice Bare Tune $Q_x$ & -- & - & 27.228 & 27.228 & 27.228 & 27.228 & 27.228 \\
Lattice Bare Tune $Q_y$ & -- & - & 25.210 & 25.210 & 25.210 & 25.210 & 25.210 \\
\hline
\multicolumn{8}{l}{\textbf{Electrons ESR}} \\
\hline
Bunch intensity & $10^{10}$ & - & 13.3 & 17.2 & 17.2 & 17.2 & 6.2 \\
Beam current & A & - & 1.93 & 2.5 & 2.5 & 2.5 & 0.225 \\
RMS normalized emittance H & $\mu$m & - & 196 & 196 & 391 & 391 & 845 \\
RMS normalized emittance V & $\mu$m & - & 34 & 18 & 26 & 26 & 71 \\
RMS emittance H & nm & - & 20 & 20 & 20 & 20 & 24 \\
RMS emittance V & nm & - & 3.5 & 1.8 & 1.3 & 1.3 & 2.0 \\
RMS bunch length $\sigma_l^*$ & mm & - & 7.0 & 7.0 & 7.0 & 7.0 & 9.0 \\
RMS $\delta_p/p$ & $10^{-3}$ & - & 0.68 & 0.68 & 0.58 & 0.58 & 1.09 \\
Collision $\beta^*$ H & m & - & 1.96 & 0.78 & 0.61 & 0.55 & 0.70 \\
Collision $\beta^*$ V & m & - & 0.21 & 0.12 & 0.071 & 0.065 & 0.057 \\
Beam-beam parameter $\xi_H$ & -- & - & 0.0569 & 0.0993 & 0.0468 & 0.0883 & 0.1086 \\
Beam-beam parameter $\xi_V$ & -- & - & 0.0434 & 0.1711 & 0.0604 & 0.1170 & 0.0995 \\
Maximum space charge $\Delta\nu_y$ & -- & - & -0.0010 & -0.0024 & -0.0004 & -0.0004 & -- \\
Laslett Tune Shift $\Delta \nu_x$ & -- & - & 0.1962 & 0.2541 & 0.1304 & 0.1304 & 0.0055 \\
Laslett Tune Shift $\Delta \nu_y$ & -- & - & -0.2245 & -0.2908 & -0.1498 & -0.1498 & -0.0060 \\
Electron beam size $\sigma_x^*$ & $\mu$m & - & 197.99 & 124.90 & 110.45 & 104.88 & 129.61 \\
Electron beam size $\sigma_y^*$ & $\mu$m & - & 27.11 & 14.70 & 9.61 & 9.19 & 10.68 \\
Lattice Bare Tune $Q_x$ & -- & - & 39.08 & 39.08 & 38.08 & 38.08 & 45.08 \\
Lattice Bare Tune $Q_y$ & -- & - & 34.14 & 34.14 & 33.14 & 33.14 & 41.14 \\
\hline
\end{tabular}
}
\caption{Consolidated EIC operational parameters and units grouped by design configuration. All listed \(\beta^*\) values are bare-lattice (before beam–beam focusing) values, and no focused or equilibrium beta function is used in this comparison. For comparison with the other energies, the Laslett tune shift for the 100 GeV case in this table is evaluated for a centered beam. For the corresponding value with an off-center beam, see Table~\ref{tb:laslett-tune-shifts-incoherent}.}
\label{tab:eic_unified_summary}
\end{table}

\subsection{\label{sec: ShortWakeField} Evaluation From Short-Range Wakefield Simulation for Single Bunch}

In the previous section, the incoherent tune shift due to indirect space-charge was evaluated from the corresponding wake function and impedance using the appropriate analytical expression for long-range wakefield. The same framework can also be applied to wake functions or impedances obtained from electromagnetic simulations of short-range wakefields for a single bunch.

The quadrupolar wake of each component in ESR and HSR can first be obtained from electromagnetic simulations using GdfidL~\cite{GdfidL, Bruns:PAC1997, Bruns:PAC1999, Bruns:EPAC2004} or ECHO3D~\cite{ECHO3D, zagorodnov2016echo}. The corresponding quadrupolar kick factor, $k_{Q,i}$, is then calculated for each component from the quadrupolar wakes of geometric origin in the time domain.
\begin{align}
k_Q &= \int_{-\infty}^{\infty} w_{Q}\lambda(s) \,ds
\end{align}

After calculating the quadrupolar kick factors for the individual vacuum components, the incoherent tune shift for a single bunch is obtained by summing the corresponding quadrupolar wake contributions using the following equation~\cite{Sakanaka2005_PRSTAB_042801}:
\begin{equation}
\begin{aligned}
\Delta\nu_{incoh} &= -\frac{q_{b}}{4 \pi (E/e)} \sum_{i} \beta_i k_{Q,i}
\end{aligned}
\end{equation}
where $k_{Q,i}$ represents the transverse quadrupolar kick factor of the corresponding component, and $q_b$ is the bunch charge.

\begin{table}[htbp]
\centering
\scriptsize
\caption{ESR wakefield simulation for a 7 mm bunch length showing contributions of individual components to the longitudinal loss factor and transverse kick factor.}
\label{tab:esr_impedance_updated}
\begin{tabular}{@{}llrrr@{}}
\toprule
\textbf{Element} &
\textbf{Label} &
\shortstack{\textbf{$\Sigma k_{\mathrm{loss}}$}\\\textbf{V/pC}} &
\shortstack{\textbf{$\Sigma k_Qy$}\\\textbf{V/pC/m}} &
\shortstack{\textbf{$\Sigma k_y$}\\\textbf{V/pC/m}} \\
\midrule
Bellows (80 mm $\times$ 36 mm)         & BLW      & 2.59e-01 & 4.16e+01 & 8.07e+01 \\
Beam Position Monitor (80 mm)          & BPM      & 4.32e-01 & 5.92e+01 & 1.01e+02 \\
Pump Port (80 mm)                      & PMP      & 2.16e-02 & 1.69e+00 & 7.66e+00 \\
Photon Absorber (69 mm gap)            & PA69     & 1.26e-01 & 3.80e+01 & 3.02e-01 \\
Flange Step (80 mm)                    & FLNG80   & 4.68e-03 & 1.79e+00 & 4.05e+00 \\
Gate Valve (80 mm)                     & GV       & 1.02e-01 & 4.69e+00 & 8.12e+00 \\
Flange Step (60 mm)                    & FLNG60   & 1.18e-03 & 9.02e-02 & 1.85e-01 \\
Photon Absorber (69 mm) with Pump      & PA69PUMP & 3.91e-02 & 7.46e+00 & 7.84e+00 \\
Photon Absorber (49 mm gap)            & PA49     & 2.89e-02 & 7.84e+00 & 3.15e-02 \\
Photon Absorber (65 mm gap)            & PA65     & 1.08e-02 & 3.89e+00 & 2.29e-02 \\
Horizontal Collimator (7.5 mm gap)     & HCLM     & 6.37e-03 & 1.52e+01 & 25.83    \\
Strip Line Kicker                      & SLK      & 4.5e-01  & 9.16e+01 & 11.14    \\
\bottomrule
\end{tabular}
\end{table}

Table~\ref{tab:esr_impedance_updated} shows an illustrative subset of ESR vacuum components and their simulated longitudinal loss factors and transverse kick factors obtained using GdfidL or ECHO3D. These component-wise results will be used to evaluate the quadrupolar wake contributions, incoherent tune shifts, and potential instabilities, with the corresponding beta functions to be finalized through the lattice analysis. These are single-bunch effects, whereas A. Chao, A. Blednykh, and Y. Shobuda consider the average current and multi-bunch effects. 

Meanwhile, the simulation-based approach evaluates each component at its actual location in the storage ring, thereby accounting for the spatial distribution of the quadrupolar impedance and explicitly incorporating the local focusing and defocusing effects.

\section{\label{sec: Discussion}  Discussion and Summary}

For the incoherent tune shifts induced by beam--beam interactions and direct space charge, as discussed in the previous sections, the tune shift depends on the transverse coordinates $(x,y)$ and the longitudinal position of individual particles within the bunch, thereby producing an intrabunch tune spread; see Fig.~\ref{figs:HSR_BeamBeam} and Fig.~\ref{figs:HSR_SpaceCharge}.

The indirect (wall-induced) space-charge incoherent tune shift, $\Delta\nu^{\mathrm{ind}}$, results from the low-frequency quadrupolar resistive-wall wake accumulated over all preceding bunches and turns. It is therefore a multi-bunch effect that scales with the average beam current $I_{\mathrm{avg}}$ and is nearly uniform within a bunch and produces little or no intrabunch tune spread. For a multibunch train with an empty gap, however, the resulting incoherent tune shift can vary from bunch to bunch, producing a bunch-to-bunch tune spread along the train, or equivalently, as a function of the longitudinal bunch position $s$. This effect has been evaluated using the PyHEADTAIL code~\cite{pyheadtail} for the CERN SPS~\cite{masessole:hb2023}. 

A similar bunch-by-bunch tune-shift study will be performed for the ESR and HSR after the wakefield simulations are completed. Furthermore, the interplay among beam–beam interactions, space-charge effects, intrabeam scattering (IBS), wakefields induced by vacuum components, and synchrotron-radiation effects in the HSR and ESR can be investigated in a subsequent study.

\begin{figure}[!ht]
\centering
\includegraphics[width=\textwidth]{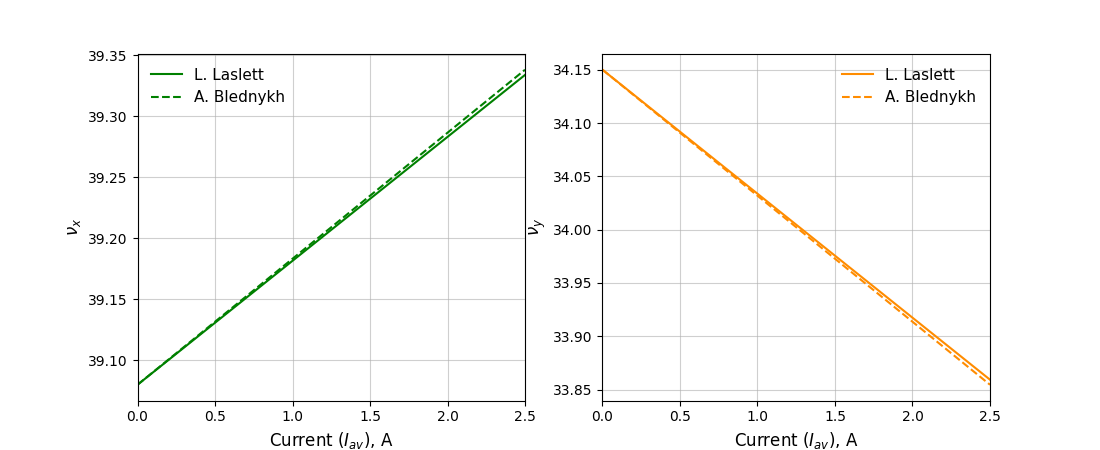}
\caption{Average-current dependence of the indirect space-charge tune shifts in the ESR at 5~GeV. The solid curves show the full-ring L. Laslett estimate, while the dashed curves show the full-ring A. Blednykh estimate. The horizontal tune increases with beam current, whereas the vertical tune decreases and approaches the integer resonance at $\nu_y=34$, motivating tune feedback correction during current accumulation. }
\label{figs:Tune_Feedback_ESR}
\end{figure}

Fig.~\ref{figs:Tune_Feedback_ESR} shows the current-dependent indirect space-charge tune shifts in the ESR at 5~GeV. The horizontal tune increases with average beam current, whereas the vertical tune decreases. The solid curves show the full-ring Laslett estimate, while the dashed curves show the estimate obtained using the Blednykh full-ring model. The two estimates agree to within $2\%$.

From Fig.~\ref{figs:Tune_Feedback_ESR}, the indirect space-charge tune shift in the ESR scales with the average beam current and represents the largest incoherent tune shift in the ESR. Based on the Laslett estimate (Tables~\ref{tab:eic_esr_5_10_18gev} and~\ref{tab:eic_unified_summary}), at the nominal current of $2.5~\mathrm{A}$, the tune shifts reach $\Delta\nu_x \approx +0.25$ and $\Delta\nu_y \approx -0.29$ at $5~\mathrm{GeV}$, and $\Delta\nu_x \approx +0.13$ and $\Delta\nu_y \approx -0.15$ at $10~\mathrm{GeV}$. The Blednykh model, with the magnetic boundary limited to the dipoles, gives $\mathrm{d}\nu_x/\mathrm{d}I = 0.0628~\mathrm{A}^{-1}$ and $\mathrm{d}\nu_y/\mathrm{d}I = -0.0719~\mathrm{A}^{-1}$ at $5~\mathrm{GeV}$. When evaluated for the full ring, the A. Blednykh model shows excellent agreement with Laslett's results (see Table~\ref{tab:tune_comparison_5GeV_ESR}).

Since these tune shifts develop progressively during injection and current accumulation, they can move the working point by more than the distance to the nearest integer resonance. For example, at $5~\mathrm{GeV}$, the vertical tune $\nu_y=34.14$ reaches the integer resonance at approximately $1.2~\mathrm{A}$. Therefore, a tune-feedback system, or feed-forward correction of the quadrupole strengths, is required in the ESR to maintain a stable working point during current accumulation.

In the HSR, the beam at injection energy is stable with the space-charge impedance alone, but becomes unstable when the geometric, resistive-wall and space-charge impedances are combined. PyHEADTAIL studies are under way to determine the octupole strengths and chromaticity required to stabilize the beam while keeping the tune spread within acceptable limits.

This work presents a unified evaluation of transverse incoherent tune shifts in the EIC Hadron and Electron Storage Rings, combining beam–beam, direct space-charge, and chamber-induced image-force effects. Analytical estimates are benchmarked against Xsuite and PyHEADTAIL simulations. 

Another important contribution is a systematic overview and comparison of theories for indirect space-charge-induced quadrupolar incoherent tune shifts, including resistive-wall impedance and classical Laslett image-force formalisms. The paper clarifies their assumptions, boundary conditions, applicability, and expected differences, and provides a practical usage guide for future EIC studies and researchers entering high-intensity beam dynamics. 

The study finds that chamber-induced shifts are modest in the HSR but can dominate the ESR at low energy, reaching approximately $+0.25$ horizontally and $-0.29$ vertically at 5 GeV and 2.5 A. These results identify tune-feedback and working-point control as essential operational requirements, and provide a practical baseline for EIC lattice optimization, impedance modeling, and collective-effects mitigation.

\section*{Acknowledgments}
This work is supported by the U.S. Department of Energy (DOE) Office of Nuclear Physics under Award Number DE-SC0023518. This work is also supported by Brookhaven Science Associates, LLC under Contract No. DE-SC0012704, and by the DOE Office of Science.

The authors would also like to acknowledge M. Blaskiewicz, Y. Luo, D. Xu, H. Lovelace for their helpful suggestions.

\bibliographystyle{unsrtnat}
\bibliography{apssamp.bib}

\section{Appendix}

\subsection{\label{sec:direct-space-charge-formula} Space-Charge Formulas}
Initially, researchers analyzed space-charge effects by modeling the beam as a continuous charged cylinder in free space. For a coasting beam, analytical expressions for the direct space-charge tune shift have been derived for several beam distributions, including a uniformly charged round beam, a uniformly charged beam with an elliptical cross section, and a round Gaussian beam. These expressions are summarized below \cite{Schindl2006_SpaceCharge}.
\begin{equation}
\label{eq:direct-space-charge-coasting}
\begin{aligned}
\Delta \nu_{xy, incoh}^{sc, direct} &= -\frac{r_0 N_{total}}{2\pi  \beta^2 \gamma^3} \frac{1}{\epsilon_{xy,geo}} && \text{Coasting Uniform Round Beam} \\[10pt]
\Delta \nu_{xy, incoh}^{sc, direct} &\cong -\frac{r_0 N_{total}}{2\pi \epsilon_{xy,geo} \beta^2\gamma^3} \frac{2}{\left(1 + \sqrt{\bar{\beta}_{yx} \epsilon_{yx,geo} / \bar{\beta}_{xy} \epsilon_{xy,geo}}\right)} && \text{Coasting Uniform Elliptic Beam} \\[10pt]
\Delta \nu_{xy, incoh}^{sc, direct} &= -\frac{r_0 N_{total}}{2\pi  \beta^2 \gamma^3} \frac{2}{\epsilon_{xy,geo}} && \text{Coasting Gaussian Round Beam}
\end{aligned}
\end{equation}
Here, $N_{\mathrm{total}}$ is the total number of particles in the ring, $r_0$ is the classical particle radius, $R$ is the machine radius, and $\beta$ and $\gamma$ are the relativistic factors. The quantity $\bar{\beta}_{x,y}$ denotes the average beta function in the horizontal and vertical planes, while $\epsilon_{x,y,\mathrm{geo}}$ is the unnormalized (geometric) emittance containing $100\%$ of the beam particles. In practice, $\epsilon_{x,y,\mathrm{geo}}$ can be expressed in terms of the rms geometric emittance or the normalized rms emittance, as shown below: 
\begin{align}
\epsilon_{xy,geo} &= 4\epsilon_{xy,geo,rms}= 4\epsilon_{xy, norm, rms}/\beta\gamma
\end{align}

For a bunched beam, following the approach described in Ref.~\cite{Reiser2008}, the direct space-charge tune shift can be derived by dividing the coasting-beam result by the bunching factor, which is defined as:

\begin{align}
B_f &= \frac{I_{\text{avg}}}{I_{\text{peak}}}
    = \frac{\sqrt{2\pi} M \sigma_z}{C},
\end{align}

where $M$ is the number of bunches in the ring, and $\sigma_z$ is the longitudinal rms bunch length. The direct space-charge tune shift for a bunched beam can then be obtained and expressed as follows:

\begin{equation}
\label{eq:direct-space-charge-bunched}
\begin{aligned}
\Delta \nu_{xy, incoh}^{sc, direct} &= -\frac{r_0 N_{bunch} }{(2\pi)^{3/2}  \sigma_z\beta^2 \gamma^3}  \frac{C}{\epsilon_{xy,geo}} && \text{Bunched Uniform Round Beam} \\[10pt]
\Delta \nu_{xy, incoh}^{sc, direct} &\cong -\frac{r_0 N_{bunch}}{(2\pi)^{3/2}  \sigma_z\beta^2 \gamma^3}   \frac{2C}{\left(1 + \sqrt{\bar{\beta}_{yx} \epsilon_{yx,geo} / \bar{\beta}_{xy} \epsilon_{xy,geo}}\right)} && \text{Bunched Uniform Elliptic Beam} \\[10pt]
\Delta \nu_{xy, incoh}^{sc, direct} &= -\frac{r_0 N_{bunch}}{(2\pi)^{3/2}  \sigma_z\beta^2 \gamma^3}  \frac{2C}{\epsilon_{xy,geo}} && \text{Bunched Gaussian Round Beam}\\
\end{aligned}
\end{equation}
where $N_{\mathrm{bunch}}$ is the number of particles in a single bunch. Eq.~\eqref{eq:direct-space-charge-bunched} can be used to estimate the direct space-charge tune shift for a bunched beam during the early design stage when the lattice is not available.

For storage rings with extremely short bunches, where the bunch length is comparable to or smaller than the transverse beam size, and with small bunch spacing, new formulas applicable to arbitrary bunch lengths and spacings for calculating the space-charge tune shift can be found in Ref.~\cite{Zhao2024_SpaceChargeTuneShift}:
\begin{equation}
\label{eq:direct-space-charge-Short}
\begin{aligned}
\Delta\nu_{xy,\text{direct}}^{\text{incoh}}
={}&
-\frac{N r_0}
{2\beta^2\gamma^3\pi^{3/2}}
\int_0^C ds\,\beta_{xy}(s)
\\
&\times
\int_0^\infty dq\,
\frac{1}
{
(2\sigma_{xy}^2+q)^{3/2}
\sqrt{
(2\sigma_{yx}^2+q)
(2\sigma_z^2+q)
}
}.
\end{aligned}
\end{equation}

\subsection{\label{sec:Off-center-Laslett} Off-Center-Beam Laslett Coefficients for a Rectangular Vacuum Chamber}
For a rectangular vacuum chamber, when the beam is displaced from the center along the horizontal axis by a fractional offset $g$ (or at a distance $gw$ from the center), the image coefficients are given by \cite{Ng2005_Book}:
\begin{equation}
\begin{aligned}
\xi_1^V=-\xi_1^H
&=
\frac{K^2(k)}{4}
\left[
\frac{k'^4 s^2 c^2}{2d^2}
-\frac{k'^2(1-2s^2)}{3}
\right.\\
&\qquad\left.
-\frac{dn^2(3-4s^2+4s^4)}{6s^2c^2}
\right],
\\[6pt]
\xi_1^V
&=
\frac{K^2(k)}{4}
\frac{k'^4s^2c^2}{d^2},
\\[6pt]
\xi_1^H
&=
\frac{K^2(k)}{4}
\left[
k'^2(1-2s^2)
+\frac{dn^2}{s^2c^2}
\right].
\end{aligned}
\end{equation}
Using the above equations, the Laslett coefficients are plotted for different aspect ratios, as shown in the figures below.

\begin{figure}[!t]
\centering
\includegraphics[width=\columnwidth]{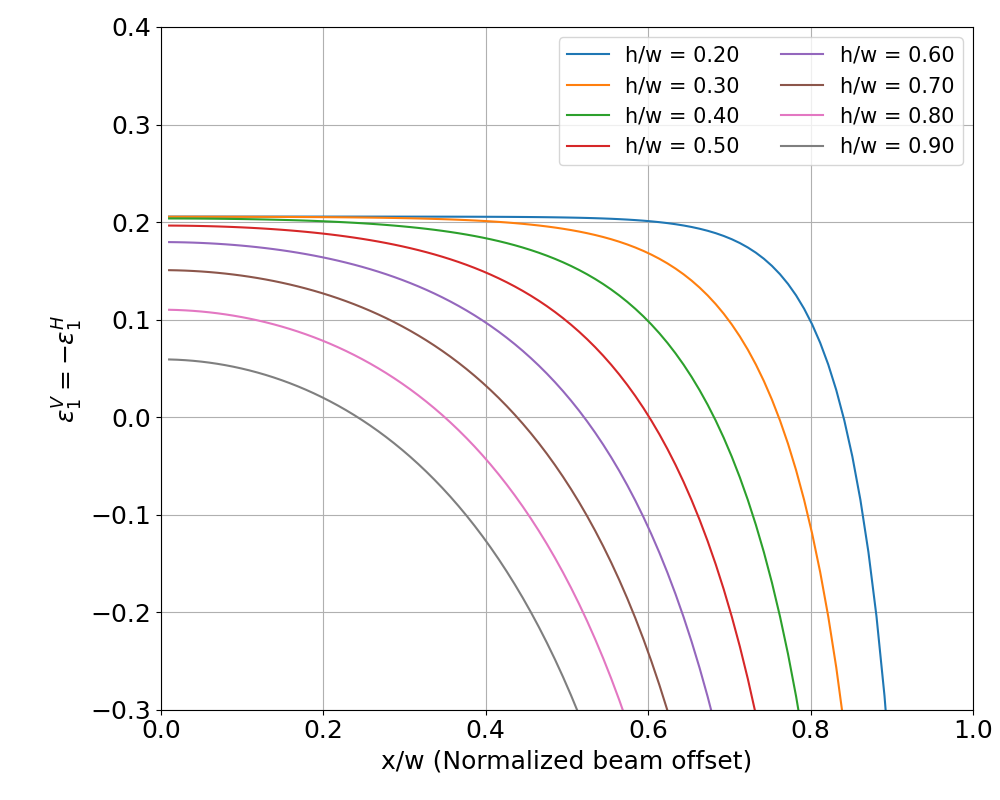}
\caption{Plot of the vertical coefficient (negative horizontal coefficient),
$\epsilon_1^V=-\epsilon_1^H$, as a function of the horizontal beam offset
$g=x/w$ from the center of a rectangular vacuum chamber for aspect ratios
$h/w=0.2,\dots,0.9$.}
\label{fig:Retangular_Lasslett_Offset_Epsilon1}
\end{figure}

\begin{figure}[!t]
\centering
\includegraphics[width=\columnwidth]{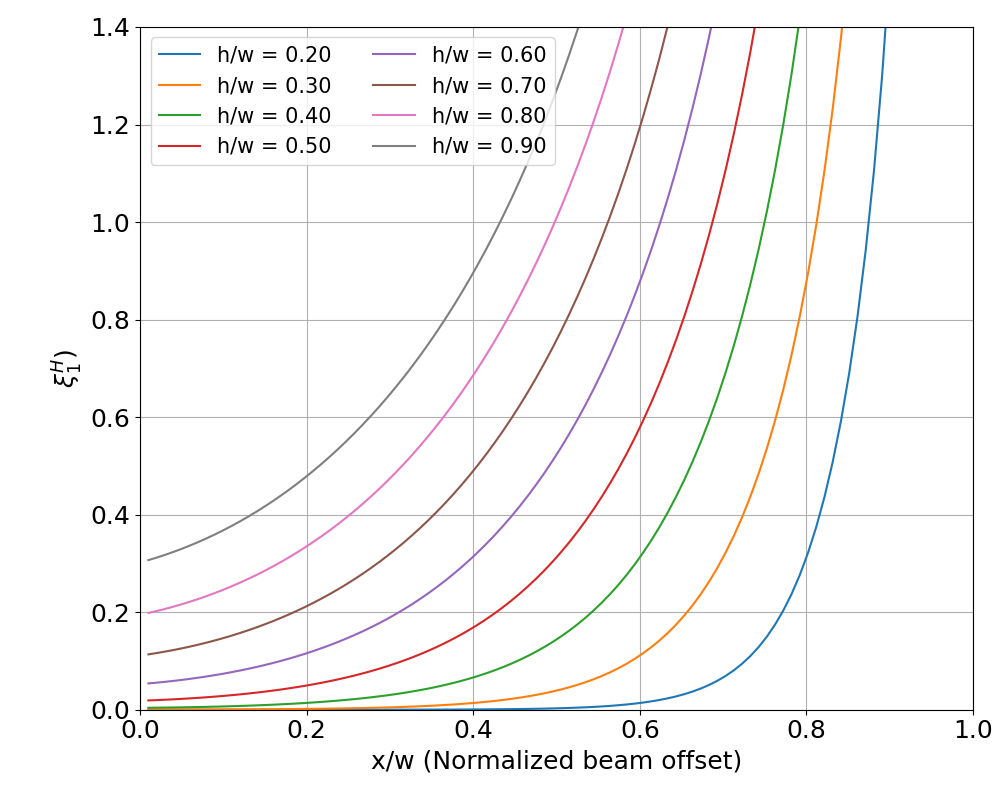}
\caption{Horizontal electric image coefficient $\xi_1^H$ as a function of the
horizontal beam offset $g=x/w$ from the center of a rectangular vacuum chamber
for aspect ratios $h/w=0.2,\dots,0.9$.}
\label{fig:Retangular_Lasslett_Offset_Xi1_H}
\end{figure}

\begin{figure}[!t]
\centering
\includegraphics[width=\columnwidth]{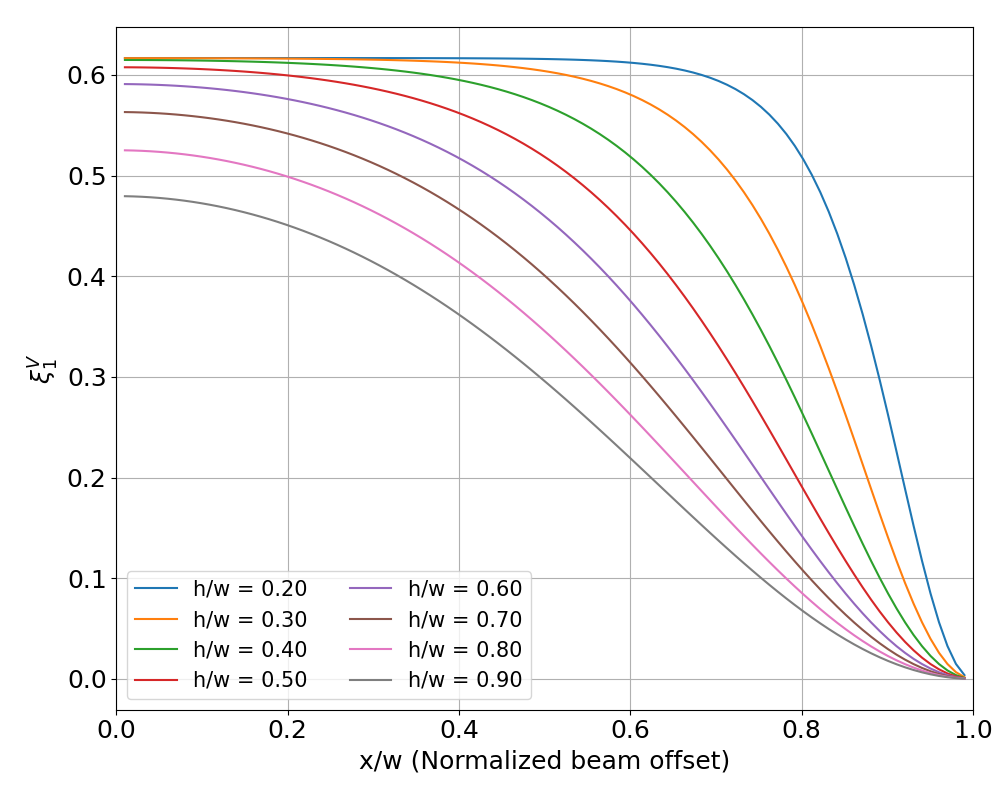}
\caption{Vertical electric image coefficient $\xi_1^V$ as a function of the
horizontal beam offset $g=x/w$ from the center of a rectangular vacuum chamber
for aspect ratios $h/w=0.2,\dots,0.9$.}
\label{fig:Retangular_Lasslett_Offset_Xi1_V}
\end{figure}

\end{document}